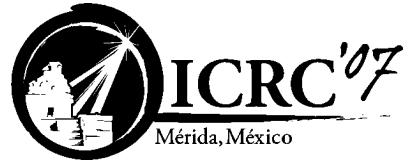

# The trigger function of the space borne gamma-ray burst telescope ECLAIRs

S. SCHANNE, B. CORDIER, D. GÖTZ, A. GROS, P. KESTENER, H. LEPROVOST, B. L'HUILLIER, M. MUR
*DAPNIA, CEA Saclay, F-91191 Gif sur Yvette, France*
*schanne @ hep.saclay.cea.fr*

**Abstract:** Gamma-ray bursts (GRB) sign energetic explosions in the Universe, occurring at cosmological distances. Multi-wavelength observations of GRB allow to study their properties and to use them as cosmological tools. In 2012 the space borne gamma-ray telescope ECLAIRs is expected to provide accurate GRB localizations on the sky in near real-time, necessary for ground-based follow-up observations. Led by CEA Saclay, France, the project is currently in its technical design phase. ECLAIRs is optimized to detect highly red-shifted GRB thanks to a 4 keV low energy threshold. A coded mask telescope with a 1024 $cm^2$ detection plane of 80×80 CdTe pixels permanently observes a 2 sr sky field. The on-board trigger detects GRB using count-rate increase monitors on multiple time-scales and cyclic images. It computes sky images in the 4-50 keV energy range by de-convolving detector plane images with the mask pattern and localizes newly detected sources with <10 arcmin accuracy. While individual GRB photons are available hours later, GRB alerts are transmitted over a VHF network within seconds to ground, in particular to robotic follow-up telescopes, which refine GRB localizations to the level needed by large spectroscopic telescopes. This paper describes the ECLAIRs concept, with emphasis on the GRB triggering scheme.

## Introduction

Gamma-ray bursts (GRB) are cosmic events, which are seen in space borne gamma-ray detectors as count rate increases during short periods of time, from tens of milli-seconds to tens of minutes. They are the signature of very energetic explosions in the Universe, mostly occurring at cosmological distances, and believed to be linked to the formation of black holes. The most popular models assume the collapse of a rotating massive star (for long duration GRB) or the merger of two neutron stars (for short GRB). Following the Gamma-ray event, afterglows are often detectable in other wavebands (visible, radio, X-rays). Those provide crucial additional information to study the physics of the event itself (constrain the GRB models, study their relativistic jets and shocks, study the link between GRB and supernovae), to determine the distance of the event (measuring the redshift of its host galaxy), to use the event as a background light source to study the foreground universe, or to use collections of those events as tools for cosmology (study the star formation history, the very first stars, constrain the cosmological parameters), or to solve questions of fundamental physics (GRB as sources of ultra-high energy cosmic rays or gravitational waves).

Currently the satellites INTEGRAL, HETE-2, and Swift deliver most of the GRB triggers to the ground-based observers, among which robotic follow-up telescopes which refine the space-given localization to a precision matching the small fields of view of the large spectroscopic 8-m class telescopes.

In 2012 the ECLAIRs gamma-ray detector, foreseen to fly on a low-earth-orbit satellite, is expected to carry on the hunt and deliver GRB triggers to the world-wide community of observers.

## Concept of ECLAIRs

The ECLAIRs project, managed by CEA Saclay, France, is currently in its detailed technical design phase, with the development phase expected to start in 2008. Unlike the previous project versions [1, 2], ECLAIRs will now be placed on an approved scientific satellite payload which is developed in collaboration between the French space agency and Chinese partners, the launch being scheduled in 2012.

The ECLAIRs flight hardware is composed of a 2D-coded mask aperture telescope (CXG, *camera for X- and gamma-rays*), with a detection plane of 80×80 CdTe pixels covering 1024 $cm^2$. It provides localizations of point sources with accuracy better than 10 arcmin in a 2 sr field of the sky in the 4 – 50 keV energy band, using a mask deconvolution technique similar to the one used in INTEGRAL. A second set of four 1D-coded mask aperture telescopes (ESXC, *soft X-ray camera*) operating from 1 – 10 keV observes the same 2 sr field in the sky, and permits to refine the localization accuracy to better than 1 arcmin for most of the GRB detected by the CXG. The low energy thresholds of its detectors render ECLAIRs more sensitive than previous missions to high-redshifted GRB, which are potentially the most interesting ones.

A real-time data link to ground is foreseen, inherited from HETE-2, with a data transfer rate of up to 600 bit/s. It uses a VHF on-board emitter and a network of about 30 VHF ground receiver stations, deployed under the satellite track. This messaging system is used among others to transmit to ground in real-time, within tens of seconds, the localization of GRB detected by ECLAIRs.

The on-board scientific processing and trigger unit (UTS) of ECLAIRs uses the CXG data in order to discover the appearance of a gamma-ray source in the sky, determines the precise localization of the source, and sends as quickly as possible (within seconds) the trigger information to the VHF network, from which it is forwarded to the observer community. For a detailed on-ground analysis, all detected photons are also stored in an on-board mass memory which is dumped to ground after a delay of up to one day via high-bandwidth X-band transceivers.

A dedicated set of ground based robotic telescopes, operating in the visible and near-infrared bands, will be used to further refine to arcsec accuracy the localization of a GRB, based on the detection of its afterglow, and to obtain a photometric redshift estimate.

The overall pointing strategy of the satellite carrying ECLAIRs is to observe the part of the sky roughly opposite to the direction of the sun (for thermal constraints), with avoidance of the galactic plane and bright X-ray sources as Sco X-1 (to reduce background). With this scheme, the Earth will be entering the field of view of the instruments every orbit, reducing the overall efficiency by about 30%. However the benefit of this strategy is that almost all detected GRB are potentially observable by large spectroscopic telescopes, because located above horizon in the Earth tropical zones harboring those telescopes.

## Functions of the scientific trigger unit

The ECLAIRs UTS (scientific trigger unit) is an on-board digital processing unit, comprising the GRB trigger and other functions. The ECLAIRs UTS electronics and software is developed at CEA Saclay, based on radiation tolerant components, among which a large Xilinx Virtex-II QPro FPGA coupled to an AT697 Leon-II processor (86 Mips, running the RTEMS operating system) and about 200 MB of radiation protected SD-RAM.

The fist function of the UTS is to receive the individual photons detected by the CXG (on 8 parallel links), as well as lists of photons detected by the ESXC, and to time-tag and send those data to the mass-memory for later X-band download.

The second UTS function is the GRB trigger, for which the CXG photons are used to detect increases in count-rates analyzed on 10 ms to 40 s timescales, followed by an attempt to localize a new source responsible for the increase (count-rate trigger) or to build cyclic sky images on 20 s to 20 min timescales in which the appearance of a new source is searched without prior count-rate increase (image trigger). For the "count-rate trigger", the UTS bins the photons received from the CXG into multiple time slices (10 ms to 40 s), energy bands and detector plane zones. The current time slices are compared to background estimators computed form past time slices. A significant count rate increase launches the source localization algorithm. In the "image trigger", the source localization algorithm is launched cyclically on predefined time scales > 20 s in order to detect slowly increasing long-duration GRB (which are more likely high-redshift GRB). For the localization of a new source on the sky using the CXG data, after a "count-rate increase trigger" or cyclically in case of the "image trigger", the UTS performs the deconvolution (by FFT) of the CXG mask pattern with the CXG detection plane detector image in order to find sources in the sky (on an extensible time-base of a few seconds to a few minutes). The raw source position (~30 arcmin accuracy) can be refined onboard by fitting the CXG point spread function to reach ~10 arcmin accuracy. Any source found is cross-checked with a catalog of know sources, using the pointing information of the satellite. In the case the detected source is new, a GRB alert is generated. The position and error box found is sent to the ESXC, in order to search within it for a position refinement down to ~1 arcmin in the so called "Vernier" mode. The GRB alert is sent by the UTS as quickly as possible (typically < 10 s after the start of the event) in an alert message to the VHF network from which it can be forwarded to the observing community. An alert is also sent to the spacecraft computer which attempts a spacecraft re-orientation maneuver (within a few minutes) in order to place the GRB in the field of view of an on-board visible-band telescope for follow-up observations. Finally the alert is also sent to the mass-memory to secure from over-writing before dump to ground the portion corresponding to the GRB data.

The third UTS function is the cyclic construction of VHF messages. In case of a gamma-ray burst trigger, the VHF message contains the alert information (CXG and SXC positions, errorbox, time, duration, type and quality of burst). GRB descriptor messages are also sent out over VHF, they can be used to compute a pseudo-redshift estimator deduced from prompt information only. They comprise in particular CXG light curves spanning over a GRB trigger period, sampled in different energy bands and with time bins of 80 ms as well as 1.28 s. During out-of-burst periods, VHF messages are also sent, in order to provide real-time housekeeping information (status and overall counting rates), as well as repetitions of previous GRB alerts.

The fourth function of the UTS is to provide house-keeping information to be down-linked over S-band a few times per day, in order to monitor the trigger process and possibly including GRB data in emergency mode.

## Algorithms of the scientific trigger

In the input stage of the UTS, photons from the CXG are received individually and accumulated in time-tagged packets sent to the mass-memory. In parallel they are energy-calibrated (into up to 8 energy bands), and stored in a photon ring-memory (of several hundreds of seconds of accumulation time). Each photon is also added into several counters, accumulating on 10 ms time-slices. There is one such counter per energy band and per detector plane zone (the zones cover all detector, two vertical and two horizontal halves, four quadrants). From the 10 ms counters, counts on timescales of 20 ms up to 80 ms are deduced by successive addition.

On these timescales, short trigger candidates are obtained by searching for an increase of the corresponding counter values over a background estimate obtained by integration of past counter values on longer timescales. In practice the background is supposed to be constant on timescales of the order of a second, such that it is sufficient to search for the maximum value of all short-time counters over this timescale and check if this one exceeds with acceptable significance the background estimate.

For triggers on timescales from 80 ms up to 40.96 s, the 80 ms counter values are used to build the integral count history. In this circular ring buffer (with several tens of minutes of history), each 80 ms time-slice contains the addition of the previous slice value with the counts detected during the current slice. Inspired from Swift [3], this trick permits by subtraction of the values of any 2 time-slices to obtain the integral of the counts detected between those moments in time. Of course, in case of roll-over of the counters, the difference will be negative, which is simply corrected for by adding the range of the counter. Such integral count histories are built for each energy band and detector zone, and each is used for triggering. Background models are built by fitting long term trends in this counting history with either linear or quadratic functions. The background estimate on timescales from 160 ms to 40.96 s is then obtained by extrapolation (or interpolation) of those functions. Any significant increase over background in such a time-slice is then a trigger candidate. The significance is obtained by comparing the excess over background to the sum of the variances of the background, the model and a minimal variance to protect against low counts. At each step in time the maximum significance is determined from each analyzed timescale, energy band and detector zone. Studies are ongoing to determine if a wavelet filter could also be used to detect significant excesses. Photons from the first significant excess over background are extracted from the photon buffer to build the corresponding detector plane image ("shadowgram"), which is in turn processed (by mask pattern deconvolution based on FFT) to search for the presence of a new source. After a successful new source localization (which takes a few seconds on the hardware considered), a first alert is sent out on the VHF. In the meantime, other significant excesses may have been found; the corresponding photons are in turn extracted form the photon buffer to build the shadowgram, for another source localization, which could result in a more precise source localization, sent out a few seconds later on the VHF.

For the "image trigger", operating on timescales exceeding 20.48 s, shadowgrams are built and sent to deconvolution cyclically every 20.48 s. The resulting sky images are searched for new sources. Higher timescales up to a few tens of minutes are obtained by summing up those sky images (count and variance images). Due to the possible presence of the Earth inside the field of view of the instrument, studies are ongoing to determine if the Earth direction has to used to model and subtract the induced background pattern prior to deconvolution.

A GRB alert is sent to the VHF in case a new source (not present in the catalog) has been found in one of the deconvolved sky images. The position of the maximum in those images is refined by fitting the point spread function of the instrument; simulations show that a localization accuracy of 5 acmin is achieved with the CXG for a 10 s burst on axis with a flux of 1 ph/cm$^2$/s. Additionally to the GRB position, the VHF alerts contain detection significances and other trigger conditions (duration, triggered energy band and detector zone, etc). After the alert has been sent out, it is repeated in case the VHF ground station did not get it properly. It is also enhanced with the ESXC source localization refinement. Those refined alerts are interleaved with complementary data for on-ground post-processing, as well as the GRB light curves in several energy bands and on 80 ms and 1.28 s timescales. The corresponding data are extracted from the count history from 1 min before to several min after the first trigger. With a 600 bit/s VHF downlink, the GRB light-curves sent to ground catch up with real-time within a few minutes after the burst, possibly before the prompt burst phase is over in case of a very long duration burst.

The previously exposed algorithms are currently under test in a so called "software model" of the UTS, before transposition to the target hardware.

## Conclusions

The ECLAIRs GRB telescope has been exposed with an emphasis on its on-board scientific processing and triggering unit, the UTS. ECLAIRs is foreseen to be mounted on a satellite scheduled for launch in 2012, and will catch up with the delivery of GRB triggers at a time when possibly no other spacecraft delivering accurate GRB localizations is in orbit anymore. Furthermore, thanks to its low energy threshold, ECLAIRs is optimized for the detection of soft GRB, X-ray flashes, and high redshift GRB, which are particularly interesting for the upcoming large ground-based quickly-reacting spectroscopic telescopes.

## References

[1] S. Schanne, J.-L. Atteia, D. Barret, S. Basa, M. Boer, B. Cordier, F. Daigne, A. Ealet, P. Goldoni, A. Klotz, O. Limousin, P. Mandrou, R. Mochkovitch, S. Paltani, J. Paul, P. Petitjean, R. Pons, and G. Skinner. *"The ECLAIRs micro-satellite for multi-wavelength studies of gamma-ray burst prompt emission"*, Trans. Nucl. Sci., vol 52, no 6, p. 2778 (IEEE Nuclear Science Symposium, October 2004).

[2] S. Schanne, J.-L. Atteia, D. Barret, S. Basa, M. Boer, F. Casse, B. Cordier, F. Daigne, A. Klotz, O. Limousin, R. Manchanda, P. Mandrou, S. Mereghetti, R. Mochkovitch, S. Paltani, J. Paul, P. Petitjean, R. Pons, G. Ricker and G. Skinner. *"The ECLAIRs micro-satellite mission for gamma-ray burst multi-wavelength observations"*, Nucl. Instrum. Meth. A567 p. 327 (2005).

[3] E. Fenimore, D. Palmer, M. Galassi, T. Tavenner, S. Barthelmy, N. Gehrels, A. Parsons and J. Tueller. *"The Trigger Algorithm for the Burst Alert Telescope on Swift"*, AIP Conference Proceedings, vol 662, p. 491 (2003).